# Tricarbon: two novel ultra-hard metallic carbon allotropes from first-principle calculations.

Samir F. Matar,[1,*] Jean Etourneau,[2] and Vladimir L. Solozhenko[3]

[1] Lebanese German University (LGU), Sahel Alma, Jounieh, Lebanon
https://orcid.org/0000-0001-5419-358X

[2] ICMCB-CNRS, University of Bordeaux, 33600 Pessac, France

[3] LSPM–CNRS, Université Sorbonne Paris Nord, 93430 Villetaneuse, France
https://orcid.org/0000-0002-0881-9761

[*] *Former DR1-CNRS senior researcher at the University of Bordeaux, ICMCB-CNRS, France*
Corresponding author email: s.matar@lgu.edu.lb and abouliess@gmail.com

*Abstract*

*Based on crystal chemistry considerations and quantum density functional theory ground state calculations, rhombohedral rh-$C_3$ and hexagonal h-$C_6$ carbon allotropes are proposed and energetically calculated as new stable ultra-hard phases likewise lonsdaleite. Along the two kinds of carbon in linear C2-C1-C2 lattice, distorted tetrahedra C2($sp^3$) with an angle of 106.17° (smaller than ideal 109.4°) and C1(sp)–like hybridizations are inferred from charge density projections. The calculated elastic constants point to a strong anisotropy of mechanical properties of rh-$C_3$ and h-$C_6$ with an exceptionally large $C_{33}$ values (1636 GPa and 1610 GPa, respectively), exceeding that of lonsdaleite (1380 GPa), due to the presence of aligned tricarbon units along the hexagonal c-axis. Both phases are characterized by large bulk moduli and high hardness values that are slightly less than those of lonsdaleite and diamond. Weak metallic behavior of both new phases is identified from electronic band structure calculations.*

**Keywords:** Tricarbon; DFT; hardness; anisotropic properties.

**Introduction with crystal chemistry considerations**

Linear triatomic arrangements are known in simple molecules like carbon dioxide. $CO_2$ in the solid state is linear in all its four polymorphs [1]. Also, triatomic linear anionic species are found in rhombohedral sodium azide $rh$-$NaN_3$ ($R$-$3m$ space group N°166) [2]. Its expression as Na{N2–N1–N2} with d(N1–N2) = 1.16 Å highlights two different nitrogen atoms: N1 and N2 belonging to two different Wyckoff positions. The structure in hexagonal setting is shown in Fig. 1a where two successive N2–N1–N2 species are isolated from each other with d{$(N_3)$-$(N_3)$} = 3.4 Å and ∠Na-N2-Na = 92.8°. Because of this "$N_3$" motifs cannot build a linear triatomic extended network which is only stabilized by the ionic interaction of $Na^+$ with $(N_3)^-$.

Oppositely, on the covalent networks side, linear arrangement of tricarbon is known within rhombohedral $B_4C$ (also expressed as $h$-$B_{36}C_9$) which is widely used as abrasive (cf. a review on rhombohedral boron embedding foreign also elements [3]). Fig. 1b depicting the large cell, highlights three C2–C1–C2 units made of two kinds of linear arrangements of carbon atoms which can be viewed as embedded in a network of three $B_{12}$ icosahedra with two kinds of boron, B1 and B2. Each C2 is in the neighborhood of three B1 forming "3B1-C2–C1–C2-3B1" triangular-base antiprism with d(B1-C2) = 1.64 Å and ∠B1-C2-B1 = 116.7°, and d(C1–C2) = 1.39 Å. Note that the angle magnitude is intermediate between $sp^2$ ∠120° in graphite, here considering rhombohedral $3R$ $C_2$ (Fig. 1c) [4], and $sp^3$ ∠109.45° such as in diamond and lonsdaleite (hexagonal diamond). Although in $NaN_3$ the triangular base pyramid N2–$Na_3$ (∠Na-N2-Na = 92.8°) shows some resemblance with C2–$B1_3$ half prism, a major difference between the two is that the rather covalent-like C-B bonding contributes efficiently to the formation of 3D covalent network providing superior mechanical properties of $rh$-$B_4C$. Tricarbon alignment is also observed in carbon suboxide $C_3O_2$ with linear O–C–C–C–O stacked in an orthorhombic structure depicted in Fig. 1d and characterized by short d(C-C) = 1.25 Å [5].

Tricarbon $C_3$ molecule is of rare occurrence, it was identified in interstellar space mainly in tails of comets [6]. The study of the molecule gained interest in recent spectroscopic works reporting on its geometry together with that of carbon dioxide $CO_2$ to address the question of linearity. Large bending vibrations amplitudes led to propose slightly bent molecules in the ground state with smaller deviation from 180° for $CO_2$ [7].

Regarding the electronic structure, the valence electron count (VEC) in $C_3$ amounts to 12, i.e. four electrons-short of $C_4$ pertaining to two dimensional (2D) $2H$ graphite or 3D lonsdaleite (Fig. 1e) with VEC = 16. Consequently the carbon hybridization in $C_3$ is not expected to be similar to that of diamond ($sp^3$) but a rather mixed hybrid $C(sp^1)$ at central carbon C1 and $C(sp^3)$–like at end C2 could be expected from the inspection of the C2–C1–C2 alignment with an expected metallic character. In this context we point out to a relatively recent work [8] of a novel stable metallic carbon allotrope i.e. $C_{18}$ (termed *H*18 carbon) in hexagonal space group *P*6/*mmm* (N°191), characterized by a mixed $sp^2$-$sp^3$ carbon bonding network was claimed from first principles investigation. Note that the metallic behavior resembles the property found herein for tricarbon because $C_{18}$ is not a multiple of 4 number of carbon valence electrons as it is the case of $C_4$ stoichiometry of graphite and lonsdaleite [9].

To the best of our knowledge there are no solid state crystal propositions available for tricarbon. Following the contextual introduction, a crystal chemistry rationale allows considering tricarbon as 3D chemical system with covalent behavior within two model approaches:

(i) rhombohedral $C_3$, (*rh*-$C_3$) based on the $NaN_3$ structure,

(ii) and hexagonal $C_6$ (*h*-$C_6$), based on lonsdaleite structure through inserting one extra carbon between two carbon atoms along *c*-axis at 2*d* (1/3, 2/3, ¼) Wyckoff position

The proposed *rh*-$C_3$ crystal structure represented using hexagonal axes setting as *h*-$C_9$ is shown in Fig. 2a. The other candidate, *h*-$C_6$ derived from lonsdaleite (Fig. 1e) is represented with a 2×2×1 supercell in Fig. 2b. Both structures show the C2-C1-C2 alignment (carbon at two distinct crystallographic sites) along *c*-axis. As demonstrated herebelow they are found to possess exceptional mechanical properties with high anisotropy along *c*-axis, mainly featured by an exceptionally high value of $C_{33}$ elastic constant.

1. **Computational framework**

The search for the ground state structure and energy was carried out using calculations based on the quantum Density Functional Theory (DFT) [10]. The plane-wave Vienna Ab initio Simulation Package (VASP) code [11, 12] was used with the projector augmented wave (PAW) method [12, 13] for the atomic potentials with all valence states especially in regard of such light element as carbon. The exchange-correlation (XC) effects within DFT were considered with the generalized gradient approximation (GGA) [14]. This XC scheme was preferred to local density approximation (LDA) [15] due to its over-binding character. The conjugate-gradient algorithm [16] was used in this computational scheme to relax the atoms onto the ground state. The tetrahedron method with Blöchl *et al.* corrections [17] and Methfessel-Paxton scheme [18] was applied for both geometry relaxation and total energy calculations. Brillouin-zone (BZ) integrals were approximated using a special **k**-point sampling of Monkhorst and Pack [19]. The optimization of the structural parameters was performed until the forces on the atoms were less than 0.02 eV/Å and all stress components below 0.003 eV/Å$^3$. The calculations were converged at an energy cut-off of 500 eV for the plane-wave basis set concerning the **k**-point integration in the Brillouin zone, with a starting mesh of 6×6×6 up to 12×12×12 for best convergence and relaxation to zero strains. In the post-treatment process of the ground state electronic structures, the total charge density and the electronic band structures are computed and visualized.

## 2. Calculations and results

A- Trends of atom resolved total energies and crystal structures

In a first step the different carbon stoichiometries were examined for their total energies from unconstrained parameter-free calculations along with successive self-consistent cycles at an increasing number of **k**-points. The results are given in Table 1 for the total energies and the atom averaged values to establish confrontations. Table 2 provides experimental and (calculated) crystal structures parameters. The total energy averaged per atom $E_{tot}$ is -9.22 eV/at. The following two lines results are relevant to new tricarbon $rh$-$C_3$ with $E_{tot}$ = -8.14 eV/at. The extended hexagonal lattice analogue $h$-$C_9$ has closely the same energy with $E_{tot}$ = -8.15 eV/at. as well as lonsdaleite-derived, $h$-$C_6$, with a similar magnitude of energy.

Regarding the crystal structures parameters, the first line in Table 2 shows calculation results for rhombohedral *3R* graphite $rh$-$C_2$ [4] illustrating good agreement between experiment and theory. The results for the two new tricarbon phases are shown in Tables 2b and 2c. The structure obtained after full geometry relaxation using sodium azide N1 and N2 atomic positions as a template (Fig. 1a) is shown in Fig. 2b, and the crystal data are presented in Table 2b. Triatomic linear C2-C1-C2 species characterized by d(C1-C2) = 1.44 Å form a 3D network connecting "$C_3$" entities together through end C2 atoms. Each C2 is within a distorted tetrahedron characterized by two angles ∠C2-C2-C2 = 106.2° and ∠C2-C2-C1 = 112.6°, respectively, slightly smaller and slightly larger than the ideal tetrahedral $sp^3$ angle of 109.4° observed for instance, in lonsdaleite (Fig. 1e).

$h$-$C_6$ crystal parameters in Table 2c show relationship with lonsdaleite, with a slight change of the $z$(C2) value from 0.0625 [9] down to 0.042 in $C_6$ due to the C1-C2 bonding resulting into two distance values for C-C: short d(C1-C2) = 1.44 Å and longer (C2-C2) = 1.55 Å, equal to the magnitude in lonsdaleite. Also due to the C3 alignment along *c*-axis, the hexagonal parameter increased from 4.12 Å [9] to 6.95 Å (Table 2c).

Further illustration of these results is obtained from the charge density projections on the atomic constituents provided in next section.

At this point the stability of linear arrangement of triatomic carbon along the hexagonal *c*-axis (Fig. 2) can be questioned in view of the multidirectional linear O–C–C–C–O stacking in carbon suboxide $C_3O_2$ (Fig. 1d) [5]. After removing the oxygen atoms, a full relaxation of the resulting $C_{24}$ orthorhombic model structure was undertaken with increasing **k**-mesh BZ integration. The total energy given in Table 1 before last line was then averaged as per one C atom, leading to E = -7.51 eV/at., i.e. with an energy increase with respect *rh*-$C_3$ or its extended *h*-$C_9$.

Lastly, we calculated the energy of a proposed novel metallic carbon allotrope *h*-$C_{18}$ [8] briefly discussed in the introduction. The calculated value $E_{tot}$ = -8.47 eV/at. is better than that of *rh*-$C_3$. Such observation is due to the close relationship with the layered-like structure at $z = 0$ shown in Fig. 3e but remains within range of the energies observed for the new tricarbon phases.

B- Charge density analysis.

To further assess the electronic and crystal structure relationship, the charge density projections onto the chemical constituents as situated in the crystal lattice are needed. The charge density output file resulting from the self-consistent calculations is analyzed with VESTA software [20]. The charge densities are obtained in yellow 3D volumes connecting atoms and showing red slices at the intersection of a crystal plane.

The charge density in 3*R* graphite expressed as *h*-$C_6$ is shown in Fig. 3a with layers of planes perpendicular to the hexagonal *c*-axis, characteristic of $sp^2$ hybridization with ∠C-C-C = 120°. A more continuous charge density joining all vertical $C_3$ units is observed in Fig. 3b with neighboring end C2 atoms connecting the adjacent segments. Each C2 is found in a distorted tetrahedron C1(C2)$_3$ with two different angles: ∠C1C2C2 = 112.6° and ∠C2C2C2 = 106.2° as shown in Fig. 3c. Note that the perfect tetrahedron angle of 109.54° in diamond is intermediate between these two values. The charge density within less stable orthorhombic tricarbon model is shown in Fig. 3d. The isolated "C3" segments show discontinuous aspects of charge density with a rather larger central C1 volume, oppositely to Figs. 2b and 2c. The mixed carbon orbital hybridization within C2-C1-C2 lets propose C1($sp^1$) and C2(distorted $sp^3$)–like hybridizations. For reminder, these nomenclatures are proper to molecular chemistry, and their use in the solid state is a guide for the reader. Lastly, Fig. 3e shows the charge density within $C_{18}$, confirming the claim [8] of mixed $sp^3$/$sp^2$ hybridization produced from present calculations. The larger stability with an energy close to graphite is then due to the ring–like $sp^2$-carbon between layers of $sp^3$–like carbon.

C- Mechanical properties.

*Elastic constants*

In view of the highly anisotropic structures with C-C-C alignments along hexagonal *c*-axis, the assessment of the mechanical properties is based on the elastic properties determined by performing finite distortions of the lattice and deriving the elastic constants from the strain-stress relationship. In hexagonal symmetry there are five independent elastic stiffness constants i.e. $C_{11}$, $C_{33}$, $C_{44}$, $C_{12}$, and $C_{13}$. Most compounds are encountered in polycrystalline form where one may consider the single crystalline grains randomly oriented. Consequently on a large scale, such materials can be considered as statistically isotropic. They are then fully described by the bulk modulus $B$ and the shear modulus $G$ that may be obtained by averaging the single-crystal elastic constants using, for instance, Voigt's method [21] based on a uniform strain.

The calculated set of elastic constants (in GPa) are:

$h$-$C_6$    $C_{11} = 753$;   $C_{12} = 120$;   $C_{13} = 40$;   $C_{33} = 1610$;   $C_{44} = 358$.

$h$-$C_9$    $C_{11} = 746$;   $C_{12} = 123$;   $C_{13} = 45$;   $C_{33} = 1636$;   $C_{44} = 356$.

All $C_{ij}$ values are positive and their combinations: $C_{11} > C_{12}$, $C_{11}C_{33} > C_{13}^2$ and $(C_{11}+C_{12})C_{33} > 2C_{13}^2$ obey the rules pertaining to the mechanical stability of the phase. From the Cij values the bulk ($B_V$) and shear ($G_V$) moduli following Voigt are formulated as:

$$B_V = 1/9 \ \{2(C_{11} + C_{12}) + 4C_{13} + C_{33}\}$$

$$G_V = 1/30 \ \{C_{11} + C_{12} + 2\ C_{33} - 4\ C_{13} + 12\ C_{44} + 6(C_{11} - C_{12})\}$$

The calculated values for the two tricarbon allotropes are then:

$h$-$C_6$ : $B_V = 392$ GPa; $G_V = 400$ GPa

$h$-$C_9$ : $B_V = 394$ GPa; $G_V = 402$ GPa

They will be further analyzed below for the corresponding hardness values.

*Hardness*

Vickers hardness ($H_V$) was predicted using two contemporary theoretical models of hardness i.e. thermodynamic model [23] and Lyakhov-Oganov model [24]. The thermodynamic model is based on crystal structure and thermodynamic properties, while Lyakhov-Oganov approach takes into account the strength of covalent bonding, degree of ionicity and directionality, as well as topology of the crystal structure. For estimation of fracture toughness ($K_{Ic}$) the Mazhnik-Oganov model [25] has been used. The results are summarized in Tables 3 and 4.

Table 3 presents Vickers hardness and bulk moduli calculated in the framework of the thermodynamic model of hardness ($B_0$), and Table 4 also presents other mechanical properties such as shear modulus ($G$), Young's modulus ($E$), the Poisson's ratio ($\nu$) and fracture toughness ($K_{Ic}$).

A slightly lower hardness of *h*-C$_6$ and *rh*-C$_3$ compared to diamond (both cubic and hexagonal) is observed for both models (Table 4). On the other hand, hardness of these phases is much higher than hardness of the vast majority of recently predicted carbon allotropes (C$_{14}$, C$_{16}$, C$_{24}$, C$_{36}$, etc.) [29-33].

A good agreement of bulk moduli of *h*-C$_6$ and *rh*-C$_3$ estimated using the thermodynamic model ($B_0$) and calculated from the set of elastic constants ($B_V$) is observed. For both phases $K_{Ic}$ is almost twice higher than the 2.8 MPa·m$^{½}$ value for single-crystal cubic BN [34] and close to the experimental value of diamond fracture toughness (5 MPa·m$^{½}$) [35].

In general, both *h*-C$_6$ and *rh*-C$_3$ have exceptional mechanical properties among all recently proposed carbon allotropes [29-33] and thus can be considered as prospective ultra-hard materials [36].

*Phonons*

Beside structural stability criteria obeyed by two new carbon phases through the relationships between the C$_{ij}$ elastic constants, we computed the phonons using 2×2×1 super-cells in hexagonal settings. Phonons are energy quanta of excitations. As photons, their energy is quantized due to the Planck constant h used in its reduced form ℏ (ℏ = h/2π) multiplied by the wave number ω with E = ℏω. Phonons are relevant to the lattice vibration modes ranging from the higher energy optical modes to the lower energy acoustic modes.

The calculated wave numbers are:

*h*-C$_6$ :

- Optical phonons range: 1459 > ω > 1007 cm$^{-1}$.
- Acoustic phonons range: 824 > ω > 389 cm$^{-1}$

*h*-C$_9$ :

- Optical phonons range: 1637 > ω > 1016 cm$^{-1}$.
- Acoustic phonons range: 991 > ω > 496 cm$^{-1}$.

All frequencies are real (positive), i.e. with absence of imaginary (negative) frequencies. This highlights dynamically and thermodynamically stable structures, enabling potential synthesis of tricarbon-based solid phases.

The ω ranges show systematically larger magnitudes in C$_9$ versus C$_6$, in agreement with the respective magnitudes of elastic constants, especially C$_{33}$. They are also in qualitative agreement with the magnitudes observed for diamond [37].

D- Electronic band structures.

Figure 4 shows the electronic band structures along the main directions of the first wedge in rhombohedral Brillouin zone.

In 3$R$ graphite the valence band VB extends over ~18 eV and shows 4 bands relevant to 2 s-like bands from -20 to -5 eV and 2 less extended p-like bands (-4 eV to -2 eV) corresponding to the two carbons atoms, each containing 2 electrons, the VB count sums up to 8 electrons leading to a closed shell electronic system. The VB is separated from the conduction band CB by a band gap at L point amounting to ~1.5 eV.

*rh*-C$_3$ exhibits a different electronic structure behavior. Although one can still count 4 bands in the VB, there are 2 more bands due to the third carbon, one of them is found fully within the VB but the 6$^{th}$ band crosses the Fermi level (E$_F$) and there is no more band gap at L. The electronic system is now a weak metal due to the crossing of E$_F$ by a dispersed (not flat) *s,p*-like band. The same result is observed for *h*-C$_6$ with the band dispersion shown in the third panel. Expectedly there are now twice more bands crossing the Fermi level than in *rh*-C$_3$.

## 3- Conclusions

New rhombohedral tricarbon is proposed based on energy and energy-derived qualities calculated within DFT. The results are backed with illustrations of the charge densities and the electronic band structures. Mechanical stabilities of both *rh*-C$_3$ (*h*-C$_9$) and *h*-C$_6$ were provided based on the sets of elastic constants and phonon wave numbers. Both new carbon allotropes are identified as highly anisotropic ultra-hard phases with metallic character.

# References

[1] Y. Han, J. Liu, L. Huang, X. He, J. Li, Predicting the phase diagram of solid carbon dioxide at high pressure from first principles. *npj Quantum Mater.* **4** (2019) 10.

[2] E.D. Stevens, H.A. Hope, A study of the electron-density distribution in sodium azide, $NaN_3$. *Acta Cryst. A* **33** (1977) 723-729.

[3] J. Etourneau, S.F. Matar, *rh*-$B_{12}$ as host of interstitial atoms: Review of a large family with illustrative study of $B_{12}\{CN_2\}$ from first-principles. *Prog. Solid State Chem.* **61** (2021) 100296

[4] H. Lipson; A.R. Stokes, A new structure of carbon. *Nature* **149** (1942) 328.

[5] A. Ellern; T. Drews; K. Seppelt. The structure of carbon suboxide, $C_3O_2$, in the solid state. *Z. Anorg. Allg. Chem.* **627** (2001) 73-76.

[6] J. Tennyson, «Molecules in Space» in *Handbook of Molecular Physics and Quantum Chemistry*, vol. 3, Chichester: John Wiley & Sons, 2003, part III, p. 358

[7] A.A. Breiera; T.F. Giesen; S.C. Ross; K.M.T. Yamada, Improved bond length determination technique for $C_3$ and other linear molecules with a large amplitude bending vibration. *J. Mol. Struct.* **1219** (2020) 128329.

[8] C.-X. Zhao, C.-Y. Niu, Z.-J. Qin, X.Y. Ren, J.-T. Wang, J.-H. Cho, Y. Jia, H18 carbon: A new metallic phase with $sp^2$-$sp^3$ hybridized bonding network. *Sci. Rep.* **6** (2016) 21879.

[9] P.D. Ownby, X. Yang, J. Liu, Calculated x-ray diffraction data for diamond polytypes. J. Amer. Ceram. Soc. **75** (1992) 1876-1883.

[10] P. Hohenberg, W. Kohn, Inhomogeneous electron gas. *Phys. Rev. B* **136** (1964) 864-871; W. Kohn, L.J. Sham, Self-consistent equations including exchange and correlation effects. *Phys. Rev. A* **140** (1965) 1133-1138.

[11] G. Kresse, J. Furthmüller, Efficient iterative schemes for ab initio total-energy calculations using a plane-wave basis set. *Phys. Rev. B* **54** (1996) 11169.

[12] G. Kresse, J. Joubert, From ultrasoft pseudopotentials to the projector augmented wave. *Phys. Rev. B* **59** (1999) 1758-1775.

[13] P.E. Blöchl, Projector augmented wave method. *Phys. Rev. B* **50** (1994) 17953-17979.

[14] J. Perdew, K. Burke, M. Ernzerhof, The Generalized Gradient Approximation made simple. *Phys. Rev. Lett.* **77** (1996) 3865-3868.

[15] D.M. Ceperley, B.J. Alder, Ground state of the electron gas by a stochastic method. *Phys. Rev. Lett.* **45** (1980) 566-568.

[16] W.H. Press, B.P. Flannery, S.A. Teukolsky, W.T. Vetterling, *Numerical Recipes*, 2nd ed. Cambridge University Press: New York, USA, 1986.

[17] P.E. Blöchl, O. Jepsen, O.K. Anderson, Improved tetrahedron method for Brillouin-zone integrations. *Phys. Rev. B* **49** (1994) 16223-16233.

[18] M. Methfessel, A.T. Paxton, High-precision sampling for Brillouin-zone integration in metals. *Phys. Rev. B* **40** (1989) 3616-3621.

[19] H.J. Monkhorst, J.D. Pack, Special k-points for Brillouin Zone integration. *Phys. Rev. B* **13** (1976) 5188-5192.

[20] K. Momma, F. Izumi, VESTA 3 for three-dimensional visualization of crystal, volumetric and morphology data. *J. Appl. Crystallogr.* **44** (2011) 1272-1276.

[21] W. Voigt, Über die Beziehung zwischen den beiden Elasticitätsconstanten isotroper Körper, *Annal. Phys.* **274** (1889) 573-587.

[22] V.A. Sachkov, V.A. Volodin, Localization of optical phonons in diamond nanocrystals. *J. Exp. Theor. Phys.* **129** (2019) 816-824.

[23] V.A. Mukhanov, O.O. Kurakevych, V.L. Solozhenko, The interrelation between hardness and compressibility of substances and their structure and thermodynamic properties. *J. Superhard Mater.*, **30** (2008) 368-378.

[24] A.O. Lyakhov, A.R. Oganov, Evolutionary search for superhard materials: Methodology and applications to forms of carbon and $TiO_2$. *Phys. Rev. B* **84** (2011) 092103.

[25] E. Mazhnik, A.R. Oganov, A model of hardness and fracture toughness of solids. *J. Appl. Phys.*, **126** (2019) 125109.

[26] P.D. Ownby, X. Yang, J. Liu, Calculated X-ray diffraction data for diamond polytypes. *J. Am. Ceram. Soc.* **75** (1992) 1876-1883.

[27] N. Bindzus, T. Straasø, N. Wahlberg, J. Becker, L. Bjerg, N. Lock, A.-C. Dippel, B.B. Iversen Experimental determination of core electron deformation in diamond. *Acta Cryst.* **A70** (2014) 39-48.

[28] V.V. Brazhkin, V.L. Solozhenko, Myths about new ultrahard phases: Why materials that are significantly superior to diamond in elastic moduli and hardness are impossible. *J. Appl. Phys.* **125** (2019) 130901.

[29] D.D. Pang, X.Q. Huang, H.Y. Xue, C. Zhang, Z.L. Lv, M.Y. Duan, Properties of a predicted tetragonal carbon allotrope: first principles study. *Diam. Relat. Mater.* **82** (2018) 50-55.

[30] W. Zhang, C. Chai, Q. Fan, Y. Song, Y. Yang, Metallic and semiconducting carbon allotropes comprising of pentalene skeletons. *Diam. Relat. Mater.* **109** (2020) 108063.

[31] X. Yang, C. Lv, S. Liu, J. Zang, J. Qin, M. Du, D. Yang, X. Li, B. Liu, C.-X. Shan, Orthorhombic $C_{14}$ carbon: A novel superhard $sp^3$ carbon allotrope. *Carbon* **156** (2020) 309-312.

[32] Q. Fan, H. Liu, L. Jiang, X. Yu, W. Zhang, S. Yun, Two orthorhombic superhard carbon allotropes: C$_{16}$ and C$_{24}$. *Diam. Relat. Mater.* **116** (2021) 108426.

[33] Q. Fan, H. Liu, R. Yang, X. Yu, W. Zhang, S. Yun, An orthorhombic superhard carbon allotrope: Pmma C24. *J. Solid State Chem.* **300** (2021) 122260.

[34] V.L. Solozhenko, S.N. Dub, N.N. Novikov, Mechanical properties of cubic BC$_2$N, a new superhard phase. *Diam. Relat. Mater.* **10** (2001) 2228-2231.

[35] N.V. Novikov, S.N. Dub, Fracture toughness of diamond single crystals. *J. Hard Mater.* **2** (1991) 3-11.

[36] V.L. Solozhenko, Y. Le Godec, A hunt for ultrahard materials. *J. Appl. Phys.* **126** (2019) 230401.

[37] R.S. Krishnan, Raman spectrum of diamond. *Nature* **155** (1945) 171.

**TABLES**

Table 1.   Energies per atom of the different carbon allotropes considered in the present work.

| Carbon allotrope | Space group | V(Å$^3$) | E$_{tot}$ (eV) | E/at. (eV) |
|---|---|---|---|---|
| $rh$-C$_2$ (*3R* graphite) [4] | *R-3m* (N°166) | 21.34 | -18.46 | -9.22 |
| $rh$-C$_3$ | *R-3m* (N°166) | 18.76 | -24.44 | -8.14 |
| $h$-C$_9$ | *R-3m* (N°166)* | 18.76 | -73.32 | -8.15 |
| $h$-C$_6$ | *P6$_3$/mmc* (N°194) | 37.15 | -48.886 | -8.148 |
| $orth$-C$_{24}$ (*C$_3$O$_2$-like*) [7] | *Pnma* (N°62) | 243.11 | -180.29 | -7.51 |
| $h$-C$_{18}$ (*H$_{18}$*) [8] | *P6/mmm* (N°191) | 113.47 | -157.48 | -8.47 |

*in hexagonal settings

Table 2. Structure parameters

a) $rh$-$C_2$ (3$R$ Graphite, $R$-$3m$, N°166) [4]. Experimental and (calculated) crystal parameters.
   $a_{hex}$ = 2.456 (2.459) Å; $c_{hex}$ = 10.41 (10.8) Å. (Fig. 1c).

| Atom | Wyckoff | x | y | z |
|---|---|---|---|---|
| C | 6c | 0.0 | 0.0 | 0.167 (0.167) |

d(C-C) = 1.419 (1.42) Å.

b) $C_3$ ($R$-$3m$, N°166) [pw] $a_{rh}$= 3.76 Å; $\alpha$ = 38.62°.

Crystal parameters in hexagonal setting, i.e. $C_9$ (Fig. 2a). $a_{hex}$ = 2.49 Å; $c_{hex}$ 10.42 Å.

| Atom | Wyckoff | x | y | z |
|---|---|---|---|---|
| C1 | 3b | 0 | 0 | ½ |
| C2 | 6c | 0 | 0 | 0.362 |

d(C1-C2) = 1.44 Å; d(C2-C2) = 1.55 Å.

c) $C_6$ ($P6_3/mmc$, N°194) [pw] (Fig. 2b). $a_{hex}$ = 2.49 Å; $c_{hex}$ = 6.95 Å.

| Atom | Wyckoff | x | y | z |
|---|---|---|---|---|
| C1 | 2d | 1/3 | 2/3 | ¼ |
| C2 | 4f | 1/3 | 2/3 | 0.042 |

d(C1-C2) = 1.44 Å; d(C2-C2) = 1.55 Å.

Table 3  Vickers hardness ($H_V$) and bulk moduli ($B_0$) of carbon allotropes calculated in the framework of the thermodynamic model of hardness [23]

|  | Space group | $a$ (Å) | $c$ (Å) | $\rho$ (g/cm$^3$) | $H_V$ (GPa) | $B_0$ (GPa) |
|---|---|---|---|---|---|---|
| $h$-C$_6$ | $P6_3/mmc$ | 2.4950 | 6.9610 | 3.1889 | 87 | 404 |
| $rh$-C$_3$ ($h$-C$_9$) | $R$-$3m$ | 2.4900 | 10.4100 | 3.2114 | 89 | 406 |
| Lonsdaleite | $P6_3/mmc$ | 2.5221[*] | 4.1186[*] | 3.5164 | 97 | 443 |
| Diamond | $Fd$-$3m$ | 3.5666[†] | – | 3.5169 | 98 | 445[‡] |

[*] Ref. 26

[†] Ref. 27

[‡] Ref. 28

Table 4  Mechanical properties of carbon allotropes: Vickers hardness ($H_V$), bulk modulus ($B$), shear modulus ($G$), Young's modulus ($E$), Poisson's ratio ($\nu$), fracture toughness ($K_{Ic}$)

| | $H_V$ | | $B$ | | $G_V$ | $E^\ddagger$ | $\nu^\ddagger$ | $K_{Ic}{}^\S$ |
|---|---|---|---|---|---|---|---|---|
| | T* | LO† | $B_0{}^*$ | $B_V$ | | | | |
| | GPa | | | | | | | MPa·m^½ |
| $h$-C$_6$ | 87 | 82 | 404 | 392 | 400 | 895 | 0.119 | 5.1 |
| $rh$-C$_3$ ($h$-C$_9$) | 89 | 83 | 406 | 394 | 402 | 900 | 0.119 | 5.1 |
| Lonsdaleite | 97 | 90 | 443 | 432 | 521 | 1115 | 0.070 | 6.2 |
| Diamond | 98 | 90 | 445** | | 530** | 1138 | 0.074 | 6.4 |

\*   Thermodynamic model [23]

†   Lyakhov-Oganov model [24]

‡   $E$ and $\nu$ values calculated using isotropic approximation

§   Mazhnik-Oganov model [25]

\** Ref. 28

# FIGURES

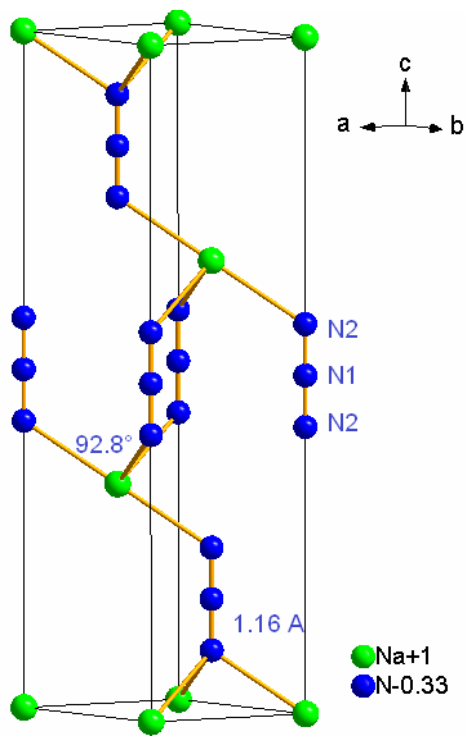

a) *rh*-NaN$_3$

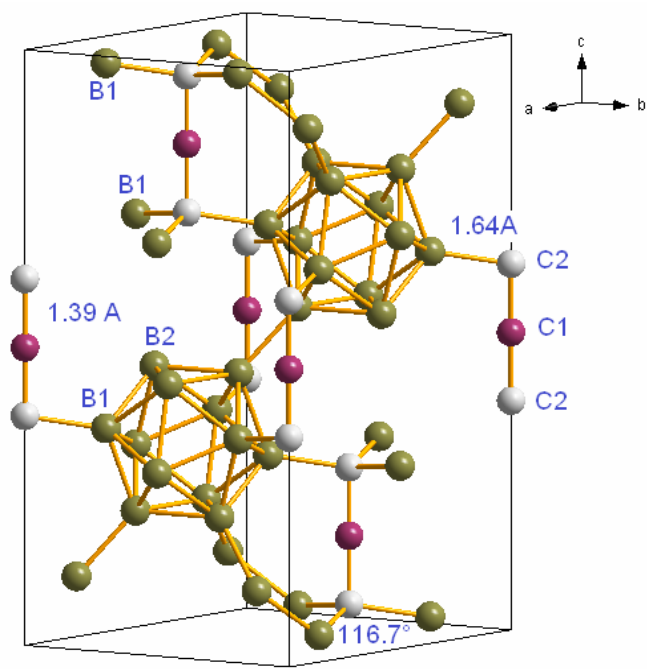

b) *h*-B$_{36}$C$_9$

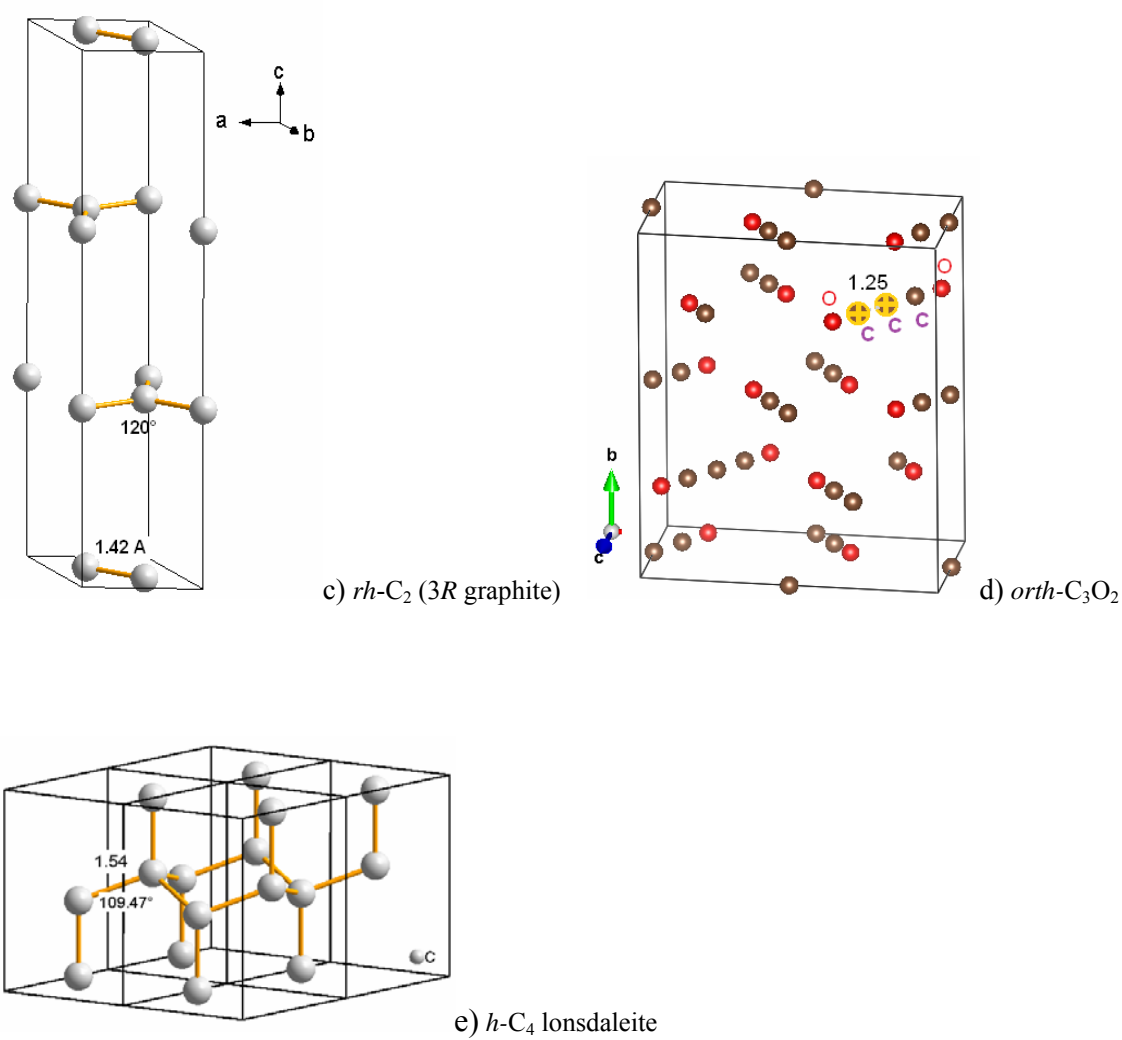

Fig. 1. Sketches of the structures a) *rh*-NaN$_3$, b) *h*-B$_{36}$C$_9$, c) *rh*-C$_2$ (3*R* graphite), d) *orth*-C$_3$O$_2$, and e) *h*-C$_4$ lonsdaleite in a 2×2×1 supercell (rhombohedral structures are represented in hexagonal settings). Remarkable interatomic distances and angles are displayed in Å and deg (°), respectively.

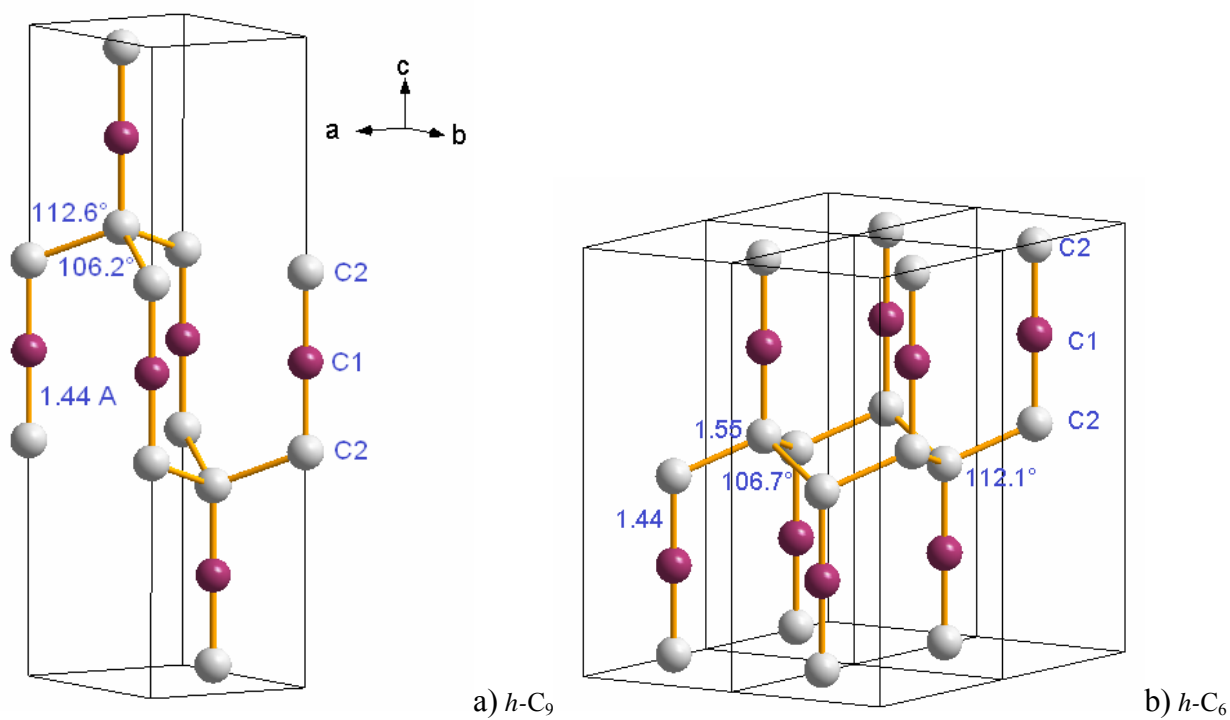

Fig. 2. Sketches of the new tricarbon rhombohedral based structures represented in hexagonal settings. a) *rh*-C$_3$ (*h*-C$_9$) and b) *h*-C$_6$ derived from C$_4$ lonsdaleite (cf. Fig. 1d) shown in a 2×2×1 supercell. Remarkable interatomic distances and angles are displayed in Å and deg (°), respectively.

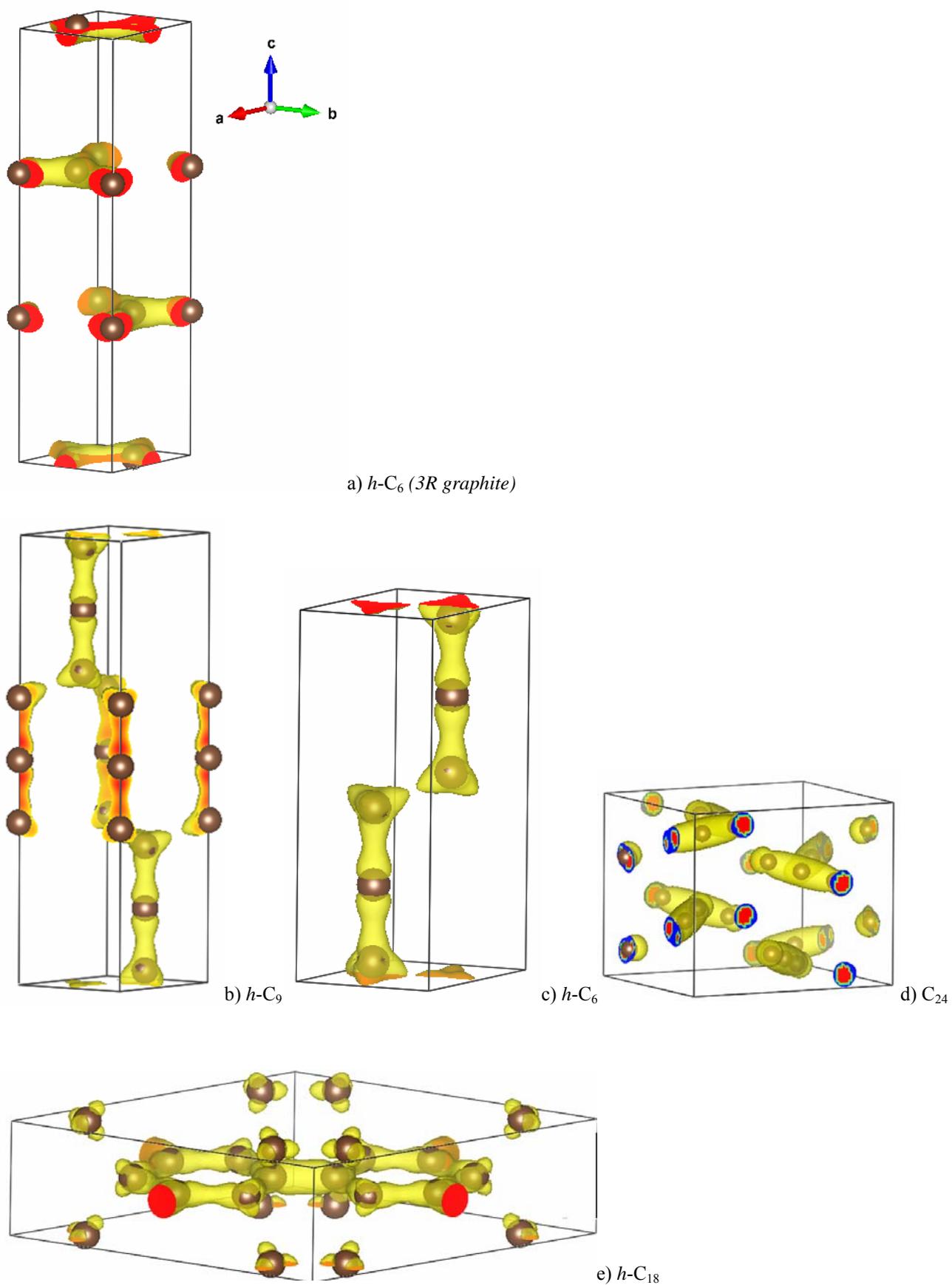

Fig. 3. Charge density projections. a) *h*-C$_6$ (hexagonal axes setting of *rh*-C$_2$); b) *h*-C$_9$ (hexagonal axes setting of *rh*-C$_3$); c) *h*-C$_6$; d) *orth.*-C$_{24}$ (from *orth*-C$_3$O$_2$); e) *h*-C$_{18}$.

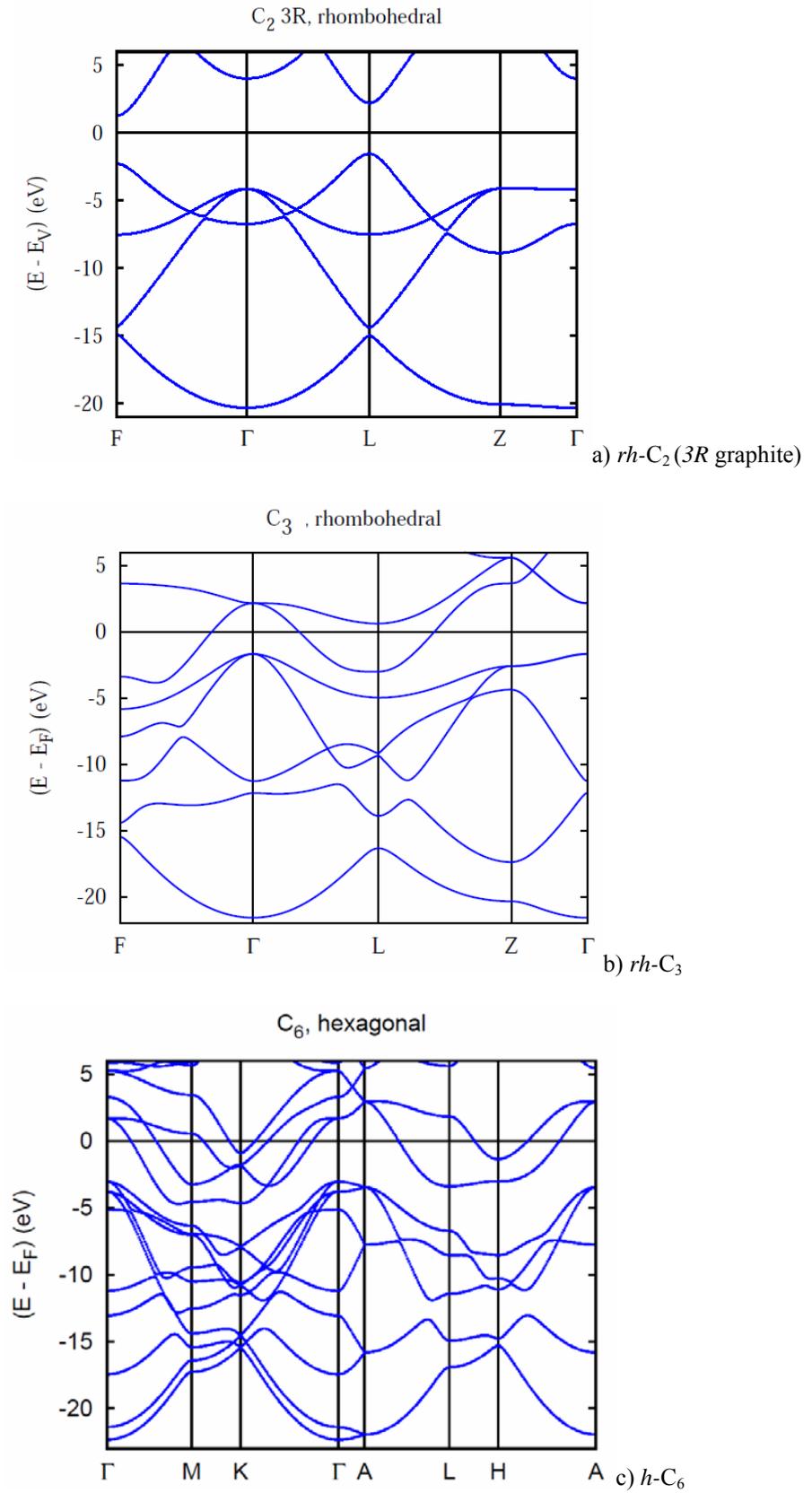

Fig. 4. Electronic band structures: a) *rh*-C$_2$ (*3R* graphite), b) *rh*-C$_3$ c) *h*-C$_6$.